\documentclass[doublecol]{epl2}
\usepackage{aas_macros}
\usepackage{amsmath}
\usepackage{amssymb}
\usepackage{graphicx}
\usepackage{hyperref}
\usepackage{physics}
\usepackage[capitalize]{cleveref}

\hypersetup{
    bookmarks=true,         
    unicode=false,          
    pdftoolbar=true,        
    pdfmenubar=true,        
    pdffitwindow=false,     
    pdfstartview={FitH},    
    pdfnewwindow=true,      
    linkcolor=red,          
    citecolor=cyan,        
    filecolor=magenta,      
    urlcolor=blue,           
    linktocpage=true
}

\newlength{\figw}
\newlength{\figh}
\newcommand{\ud}{\dd}

\newcommand{\eisss}[1]{{\scriptscriptstyle{#1}}}
\newcommand{\bdata}[1]{\boldsymbol{D}_{#1}}

\newcommand{\uS}{\mathrm{S}}
\newcommand{\uKL}{\mathrm{KL}}
\newcommand{\uSPT}{\mathrm{SPT}}
\newcommand{\uBK}{\mathrm{BK}}
\newcommand{\usssKL}{\eisss{\uKL}}
\newcommand{\calD}{\mathcal{D}}

\newcommand{\calP}{\mathcal{P}}
\newcommand{\calI}{\mathcal{I}}

\newcommand{\alphac}{\alpha}

\newcommand{\prior}[2]{\pi_{#1}\negthinspace\left(#2\right)}
\newcommand{\post}[2]{P\negthinspace\left(#1|#2\right)}

\newcommand{\BB}{B\negthinspace B}
\newcommand{\TT}{TT}
\newcommand{\EE}{E\negthinspace E}
\newcommand{\TE}{T\negthinspace E}

\newcommand{\COBAYA}{\texttt{cobaya}}
\newcommand{\CAMB}{\texttt{CAMB}}
\newcommand{\SROLL}{\texttt{SRoll2}}
\newcommand{\PLANCK}{\textit{Planck}}
\newcommand{\LOWLTT}{\texttt{lowlTT}}
\newcommand{\EBOSS}{\texttt{eBOSS}}
\newcommand{\SPTDONE}{\texttt{SPT3GD1}}
\newcommand{\CANDL}{\texttt{candl}}

\newcommand{\DKL}[1]{D_{\usssKL}^{#1}}
\newcommand{\nS}{n_{\eisss{\uS}}}

\title{BaBy Cosmic Tension}

\author{Christophe Ringeval\thanks{\email{christophe.ringeval@uclouvain.be}}}
\shortauthor{Christophe Ringeval}
\institute{Cosmology, Universe and Relativity at Louvain (CURL),
  Institute of Mathematics and Physics, University of Louvain, 2 Chemin
  du Cyclotron, 1348 Louvain-la-Neuve, Belgium}

\abstract{We show that the recently released $B$-mode polarisation
  data from the South Pole Telescope (SPT) favour a non-vanishing
  contribution of primordial gravitational waves of inflationary origin
  which is in tension with the previous BICEP-Keck (BK)
  measurements. Our analysis uses the third-order slow-roll primordial
  power spectra, with theoretically motivated priors, on the
  multifrequency SPT likelihoods complemented by the latest Planck
  satellite data products. The SPT measurements provide $1.0$ bit of
  information gain on the first slow-roll parameter, which is higher
  than the $0.9$ bit provided by BK even though the SPT sensitivity is
  five times lower. Moreover, the Bayesian dimensionality on the same
  parameter exceeds $1.5$ for SPT versus $0.3$ for BK showing that it
  is overconstrained by the SPT data. Even if this $\BB$-tension could
  be the result of a yet to be understood foreground, our findings
  should motivate for a closer analysis of this unexpected $B$-modes
  excess.}
\begin{document}

\maketitle

\section{Introduction}

Within General Relativity, the intrinsic $B$-mode polarisation of the
Cosmic Microwave Background (CMB) anisotropies can only be sourced by
gravitational effects coming from vector and tensor cosmological
perturbations~\cite{Kamionkowski:1996zd, Seljak:1996gy}. As such, the
search for $B$-modes is considered as a privileged channel for the
discovery of primordial gravitational waves and/or active sources such
as cosmic strings~\cite{Seljak:1997ii, Pogosian:2007gi,
  Ringeval:2010ca}. However, $B$-type polarisation of CMB photons can
also be generated after the last scattering surface by various
secondary effects~\cite{Fidler:2017irr}. The most prominent being a
conversion from $E$- to $B$-modes due to gravitational
lensing~\cite{Lewis:2006fu, Smith:2007rg, Planck:2018lbu}. This effect
is, however, theoretically well known, dominates at small angular
scales and can be disambiguated from primary $B$-modes by multiscale
observations and cross-correlations with other
probes~\cite{SPT-3G:2025zuh}. Other secondary sources of $B$-mode
polarisation exist, such as polarised thermal emission from dust
grains aligned with the galactic magnetic field or from ice crystals
within the atmosphere~\cite{SPT-3G:2024yab,
  Pan-ExperimentGalacticScienceGroup:2025vcd}. Contamination by these
unwanted foregrounds and nuisances can however be mitigated through
dedicated experimental set-up or by including them in the data
analysis~\cite{Martin:2014lra, Flauger:2014qra, Planck:2018gnk}.

The most sensitive measurements of the large-scales $B$-mode
polarisation have been achieved by the BICEP-Keck collaboration and
they have been used to provide an upper bound on the amount of
primordial gravitational waves~\cite{BicepKeck:2021ybl}. In terms of
the so-called tensor-to-scalar ratio, Ref.~\cite{BICEP:2021xfz}
reports $r<0.032$ ($95\%$ CL) with an uncertainty
$\sigma_r=0.014$. From a theoretical point of view, the most favoured
mechanism that would explain the existence of primordial gravitational
waves is Cosmic Inflation~\cite{Starobinsky:1980te, Guth:1980zm,
  Linde:1981mu, Mukhanov:1981xt}. In the inflationary paradigm,
cosmological perturbations arise from the quantum fluctuations of the
field-metric system during a phase of quasi-de Sitter accelerated
expansion~\cite{Mukhanov:1990me}. This mechanism generates both scalar
and tensor perturbations~\cite{Mukhanov:1985rz, Mukhanov:1988jd}. For
all slow-roll single field models, the simplest incarnations of Cosmic
Inflation, it is possible to analytically derive the functional shape
of the primordial scalar and tensor power spectra in terms of the
comoving wavenumbers $k$. As of today, these expressions are known up
to third order in the so-called Hubble-flow expansions where one
introduces a hierarchy of small slow-roll parameters
$\epsilon_i$~\cite{Schwarz:2001vv,Auclair:2022yxs}. The first of these
parameters, $\epsilon_1$, has an unknown order of magnitude and, at
leading order, gives the tensor-to-scalar ratio $r= 16 \epsilon_1 +
\order{\epsilon_i^2}$. Together with the second one, $\epsilon_2$,
they fix the spectral index of the scalar perturbations $\nS = 1 - 2
\epsilon_1 - \epsilon_2 + \order{\epsilon_i^2}$, and so forth at
higher orders. A Bayesian analysis of the BK data based on these
theoretically motivated power spectra, and complemented with other
cosmological data sets, has been presented in
Refs.~\cite{Martin:2024qnn, Ballardini:2024irx}. The authors report a
weak preference for $\log(\epsilon_1)>-3.9$ at $68\%$ confidence level
(CL) and an upper bound: $\log(\epsilon_1) < -2.6$ at $98\%$ CL. Since
$r \simeq 16\epsilon_1$, any preference for a non-vanishing
$\epsilon_1$ is a preference for non-vanishing primordial
gravitational waves of slow-roll inflationary origin. As explicitly
stated in Ref.~\cite{Martin:2024qnn}, these figures are
non-statistically significant, but they do raise the question of the
existence, and origin, of this small residual $B$-modes excess.

Earlier this year, the South Pole Telescope
collaboration~\cite{Carlstrom:2009um} has released a competitive
measurement of the CMB $B$-modes angular power spectrum in
Ref.~\cite{SPT-3G:2025vtb}. A subset of their measurements actually
includes the BK field of view in the Southern Sky and this provides a
new and independent observation of the $B$ polarisation at large
angular scales. Ref.~\cite{SPT-3G:2025vtb} reports an upper bound on
$r<0.25$ ($95\%$ CL), with an uncertainty $\sigma_r= 0.067$. These
bounds are much less constraining than the ones reported by BK and
this could be interpreted as the consequence of the current lower
sensitivity of the SPT data compared to BK (which essentially arises
from a smaller integration time). However, as we show below, when
performing data analysis from the (third-order) slow-roll primordial
power spectra, the SPT data are actually favouring a non-vanishing
contribution for primordial gravitational waves, $\log(\epsilon_1)
>-2.8$ at $68\%$ CL and $\log(\epsilon_1) >-4.5$ at $95\%$ CL. These
(weak) lower bounds actually confirm that there is a residual $B$-modes
excess at large angular scales within the sky portion probed by the
two independent telescopes. Although, taken alone, this preference is
still non-statistically significant, the SPT lower bound
$\log(\epsilon_1)>-2.8$ corresponds to domains where the BK posterior
is very small. This is quite unexpected, all the more so because the
sensitivity of SPT is lower than BK.

The paper is organised as follows. In the next section, we present our
theoretical hypothesis, priors and data sets used to fairly compare
the consequences of using either BK or SPT for performing Bayesian
inference on the slow-roll inflationary parameters. Then, we present
our results, which, among others, include the marginalised posteriors
of all the slow-roll parameters from the SPT data. The differences
between BK and SPT on the $\log(\epsilon_1)$ posterior distribution are
then discussed in terms of information gain and Bayesian
dimensionality. Finally, in the conclusion, we discuss the effects of
including other data sets and critically assess the significance and
consequences of our findings.

\section{Data analysis}
\label{sec:analysis}

Maximal inference on the shape of the slow-roll primordial power
spectra can be achieved by combining data sets which are both
sensitive to small and large angular scales and, most importantly,
which do not exhibit any inconsistencies or
tensions~\cite{Handley:2020hdp}. As our objective is to compare the
effects of incorporating either the BK or SPT $B$-mode measurements,
the chosen data sets have to be consistent with any of them. For these
reasons, we have considered the {\PLANCK} NPIPE CamSpec data
products~\cite{Rosenberg:2022sdy}(providing the CMB $\TT$, $\TE$ and
$\EE$ angular power spectra) complemented with the large-scales $\EE$
channel from the {\SROLL} maps~\cite{Delouis:2019bub}. In order to
ease lensing disambiguation and tighten possible degeneracies with the
standard cosmological parameters, we have also included the {\PLANCK}
PR4 lensing likelihoods from Refs.~\cite{Carron:2022eum,
  Carron:2022eyg} as well as the Baryon Acoustic Oscillations (BAO)
from the multiple \texttt{DR16} tracers from
{\EBOSS}~\cite{Bautista:2020ahg, Gil-Marin:2020bct, deMattia:2020fkb,
  Hou:2020rse, Neveux:2020voa, duMasdesBourboux:2020pck}. In order to
improve ``lever-arm'' constraints between large and small angular
scales, we have finally added the latest SPT measurements of the
$\TT$, $\TE$ and $\EE$ spectra at high
multipoles~\cite{SPT-3G:2025bzu} (\texttt{\SPTDONE}). In the
following, these data sets will be collectively referred to as
$\calD$. Let us mention that neither BAO from the Dark Energy
Spectroscopic Instrument (DESI)~\cite{DESI:2025zgx} nor the latest
Atacama Cosmology Telescope (ACT) data~\cite{ACT:2025fju} are included
in $\calD$ due to their reported tensions with either the standard
$\Lambda$CDM model~\cite{CosmoVerseNetwork:2025alb} or the {\PLANCK}
preferred values for the spectral
index~\cite{Ferreira:2025lrd}. Moreover, in order to fairly compare
the BK vs SPT lone constraints on the possible amount of inflationary
gravitational waves, we have not included the lowest multipoles of the
{\PLANCK} $\TT$ power spectrum (the so-called {\LOWLTT}
likelihood~\cite{Aghanim:2019ame}). Indeed, in addition to $B$-modes,
spin-$2$ cosmological perturbations have a small but non-vanishing
contribution to the temperature CMB power spectrum at large angular
scales. The inclusion of these unwanted data sets will be however
discussed in the Conclusion.

On the theory side, the late time universe is assumed to be described
by a standard spatially flat $\Lambda$CDM model (with standard
reionisation history). Cosmological perturbations of both scalar and
tensor types are assumed to be generated by a primeval phase of
slow-roll single field inflation. Both the scalar and tensor power
spectra are taken as analytical functional of the four slow-roll
parameters required at third-order:
$\calP_\zeta(k,\epsilon_1,\epsilon_2,\epsilon_3,\epsilon_4)$ for the
comoving curvature perturbation $\zeta$ and
$\calP_h(k,\epsilon_1,\epsilon_2,\epsilon_3,\epsilon_4)$ for
gravitational waves. Their explicit expressions are rather long and
can be found in Ref.~\cite{Auclair:2022yxs}, see Eqs.~(44) and (54).

Bayesian data analysis has been computationally performed by exploring
the full parameter space with Markov-Chains-Monte-Carlo (MCMC) methods as
implemented in the Python package
\href{https://github.com/CobayaSampler/cobaya}{\COBAYA}~\cite{Torrado:2020dgo}. For
this purpose, the theoretical CMB angular power spectra have been
computed with a modified version of the \href{https://github.com/cosmicinflation/camb}{\CAMB}
code~\cite{Lewis:1999bs,Howlett:2012mh}, which deals with the
aforementioned third-order slow-roll power spectra. All the
likelihoods required for this analysis are publicly available and
provided by their respective collaborations. Let us mention that, in
order to not overlook any possible degeneracies with foreground
parameters, we have used the full {\SPTDONE} \texttt{v2.0.0}
multifrequency likelihood for the SPT data (as opposed to the
so-called ``lite'' implementation) which is interfaced with the public
Python package
\href{https://github.com/SouthPoleTelescope/spt_candl_data}{\CANDL}~\cite{Balkenhol:2024sbv}. The
dimensionality of the parameter space being relatively large ($68$),
the MCMC exploration has been pushed up to a total number of collected
samples exceeding $30$ millions. For both cases, this corresponds to a
Gelman-Rubin criterion of convergence $R-1 \lesssim \order{1} \times
10^{-3}$.

The prior probability distributions have been set as flat priors for
the standard $\Lambda$CDM cosmological parameters. Concerning the
slow-roll parameters, we have chosen flat priors for
$\epsilon_{2,3,4}\in[-0.2,0.2]$ ensuring the consistency of the
expansion. The first parameter $\epsilon_1$ is a strictly positive
number, the order or magnitude of which being unknown. An
uninformative prior would therefore be a Jeffreys' prior on
$\epsilon_1$, which can also be implemented as a flat prior on its
logarithm $\log(\epsilon_1)\in[-5,-0.7]$. The lower bound is
arbitrarily set to a small undetectable threshold while the upper
bound matches the upper bound of the others $\epsilon_i$. As for the
nuisance and foreground parameters, prior distributions are
automatically set by their respective likelihood
implementations. Nonetheless, for the SPT data, we have extended the
prior on the power-law index $\alphac\in[-5.5,-1]$ of the angular
dependence of the galactic dust emission and on the power law index
$\beta\in[0,5]$ encoding its spectral distortion as to ensure a wide
enough marginalisation for our choice of data sets $\calD$.

\section{Results}
\label{sec:info}
\begin{figure}
  \begin{center}
    \includegraphics[width=0.98\figw]{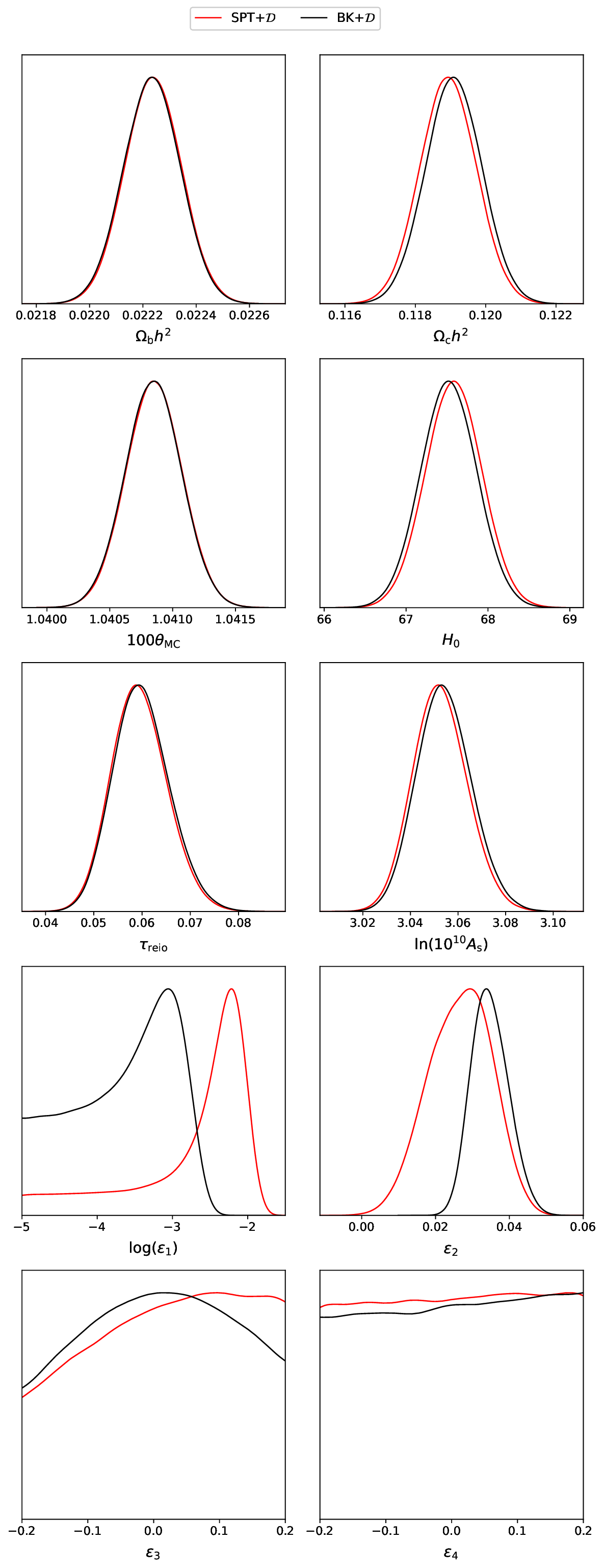}
    \caption{One-dimensional marginalised posterior distribution for
      the cosmological and slow-roll parameters obtained from either
      the SPT (red) or BK (black) $B$-mode data.}
    \label{fig:Psr1D}
  \end{center}
\end{figure}

The resulting one-dimensional marginalised posteriors for the
cosmological and slow-roll parameters are plotted in \cref{fig:Psr1D}
for BK (black curves) and SPT (red curves). Most of the posteriors are
nearly identical between the two data sets and, on Bayesian grounds,
$\uSPT+\calD$ and $\uBK+\calD$ are certainly
compatible~\cite{Martin:2014lra, SPT-3G:2025vtb}. However, we have
selected our data sets in such a way that $\log(\epsilon_1)$ is solely
constrained by $B$-modes and, as seen in \cref{fig:Psr1D}, SPT and BK
do not yield the same probability distribution. The posterior for
$\log(\epsilon_1)$ associated with the SPT $B$-mode data (red curve)
is peaked around a most probable value which is quite disfavoured by
the BK data (black curve). At the same time, one can also notice a
significant widening of the posterior on the second slow-roll
parameter $\epsilon_2$, which is due to its degeneracies with
$\epsilon_1$ in the definition of the spectral index. The two
remaining slow-roll parameters, $\epsilon_3$ and $\epsilon_4$ are
essentially unconstrained.

The shape of the $\log(\epsilon_1)$ posterior from SPT (and, to some
extend from BK) is somehow unexpected. If no excess $B$-modes were
present, one would have obtained a Heaviside-like distribution: flat
at small $\log(\epsilon_1)$ values and vanishing for the large enough
values to be within the SPT (BK) sensitivity domain. On the contrary,
here, both SPT and BK exhibit a posterior peaked at radically
different values having a long non-vanishing tail for small
$\log(\epsilon_1)$ (left part). For SPT, the peaked is more pronounced
and the posterior leads to the one- and two-sigma lower bounds already
quoted in the introduction: $\log(\epsilon_1) > -2.8$ at $68\%$ CL,
and $-4.5$ at $95\%$. At three-sigma, only an upper bound remains
$\log(\epsilon_1)<-1.8$ ($98\%$ CL) and, taken alone, this posterior
does not support any claim of detection.

Still, one can quantify how much information is carried by these
posteriors when compared to the prior distribution. Denoting $P$ the
posterior distribution and $\pi$ the prior, for $\bdata{x}=\calD+\uSPT$ or
$\bdata{x}=\calD+\uBK$, the Kullback-Leibler divergence is given
by~\cite{kullback1951}
\begin{equation}
\DKL{x} = \int \post{\log \epsilon_1}{\bdata{x}} \ln\left[
\dfrac{\post{\log \epsilon_1}{\bdata{x}}}{\prior{}{\log \epsilon_1}}
\right] \ud \log \epsilon_1,
\end{equation}
which is the first moment of $\calI=\ln(P/\pi)$, the Shannon's
entropy. One obtains, in bits,
\begin{equation}
  \DKL{\uSPT}=0.97, \qquad \DKL{\uBK} = 0.86,
\end{equation}
which is unexpected as BK has a five times higher sensitivity than
SPT. If no signal were present, $\DKL{x}$ would have been higher for
BK than SPT, by an amount directly given by the difference of
parameter space measure ruled-out by the data. This inverted hierarchy
in information gain comes from the peak of the SPT posterior which
overcomes its smaller sensitivity than BK. The Bayesian dimensionality
is the quantity of interest to assess how peaked a distribution
is~\cite{Handley:2019pqx}. It is also a measure of the number of
constrained parameters by the posterior. It reads
\begin{equation}
d_x  = 2\left(\ev{\calI^2}_x - \ev{\calI}_x^2\right),
\end{equation}
where the mean values are over the posterior,
$\post{\log\epsilon_1}{\bdata{x}}$ for our purpose. As discussed in
Ref.~\cite{Handley:2019pqx}, the Bayesian dimensionality of a flat
distribution vanishes whereas a one-dimensional Gaussian distribution
has $d=1$. From the posterior of $\log(\epsilon_1)$ plotted in
\cref{fig:Psr1D}, we obtain
\begin{equation}
  d_\uSPT=1.57, \qquad d_\uBK=0.31.
\end{equation}
As a result, the SPT posterior is, in this respect, overconstraining
and this justifies why its information gain over the prior is
surprisingly larger than BK.

\begin{figure}
  \begin{center}
    \includegraphics[width=\figw]{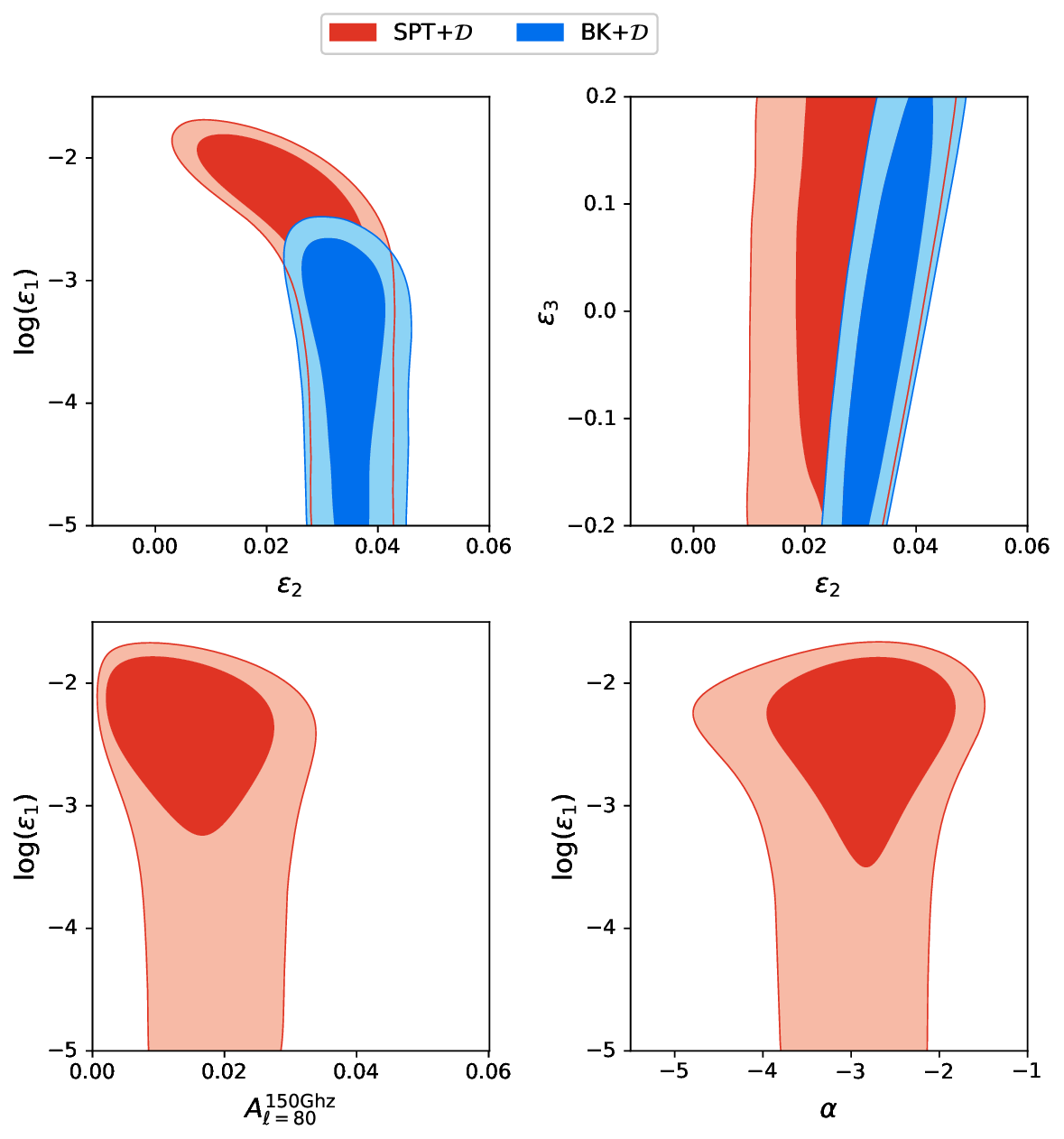}
    \caption{Two-dimensional marginalised posterior distribution for
      selected pairs of correlated parameters from the SPT (red) or BK
      (blue) $B$-mode data.}
    \label{fig:Psr2D}
  \end{center}
\end{figure}

For completeness, we have represented in \cref{fig:Psr2D} the
two-dimensional marginalised posteriors of various relevant pairs of
parameters including the amplitude and power indices of the polarised
dust foreground. The nuisance posteriors are matching the ones
reported in Ref.~\cite{SPT-3G:2025vtb}.

Let us finally recap that our results are derived from theoretically
motivated priors on the slow-roll parameters, and these are not the
same as taking flat priors on the more usual phenomenological
parameters which are the spectral index $\nS$ and the tensor-to-scalar
ratio $r$. The effect of changing priors on $r$ has been discussed in
various other works~\cite{Leach:2002dw, Hergt:2021qlh} and they have
shown that choosing a flat prior on $\log(\epsilon_1)$ actually
disfavours large values of $r$. This reinforces the fact that the peak
in the posterior of $\log(\epsilon_1)$ is completely driven by the SPT
$B$-mode data.

\begin{figure}
  \begin{center}
    \includegraphics[width=\figw]{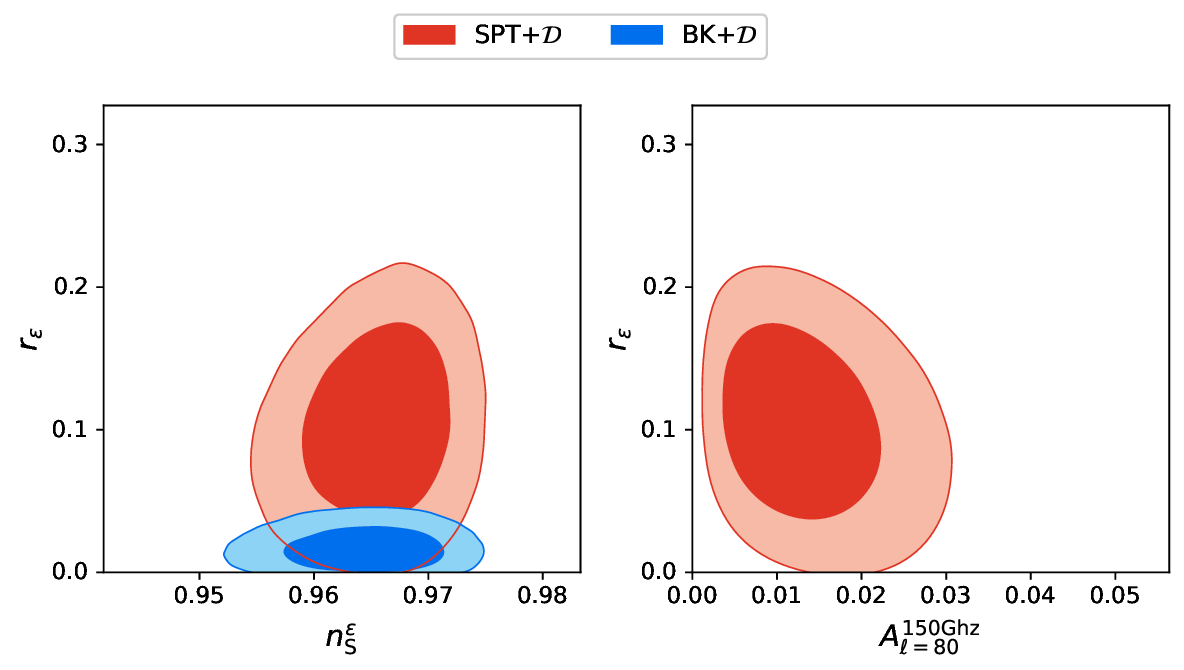}
    \caption{Two-dimensional marginalised posterior distribution of
      the \emph{derived} power law parameters $\nS^\epsilon$ and $r_\epsilon$
      obtained by importance sampling from the slow-roll posteriors
      associated with the SPT (red) or BK (blue) data (at
      third-order).}
    \label{fig:Ppl2D}
  \end{center}
\end{figure}

In more quantitative terms, $\nS$ and $r$ are not in one-to-one
correspondence with the slow-roll parameters and performing data
analysis on the $\epsilon_i$ or on $(\nS,r)$ is not
equivalent. However, at a given order in slow-roll, one has
$\nS^\epsilon=\nS(\epsilon_i)$ and $r_\epsilon=r(\epsilon_i)$ (see
Appendix C of Ref.~\cite{Auclair:2022yxs} for an explicit expression
of these functions). The posteriors for the derived parameters
$(\nS^\epsilon,r_\epsilon)$ can therefore be computed by importance
sampling from the ones of the $\epsilon_i$ and they have been
represented in \cref{fig:Ppl2D}. The two-dimensional posteriors
involving $r_\epsilon$ for SPT are quite shifted compared to BK, while
the two-sigma contours are nearly closed around a non-vanishing value
of $r_\epsilon\simeq 0.1$. This maximum probability value for
$r_\epsilon$ is actually close to the one found in
Ref.~\cite{SPT-3G:2025vtb} for the pure power-law parameter $r$, and,
up to the more constraining contours obtained here, both analysis are
compatible. Let us notice that the posterior spread on
$\nS^{\epsilon}$ is mostly driven by the $TT$, $TE$ and $EE$ data and
is essentially uncorrelated to $r_{\epsilon}$.

\section{Conclusion}
\label{sec:conc}

We find that the recent SPT measurements of the CMB $B$-modes within
the same region as BK favours, at two-sigma, non-vanishing primordial
gravitational waves of slow-roll inflationary origin. Although this is
non-statistically significant, we have shown that this result is in
tension with the previous BK measurements using various measures of
the amount of information contained in the posteriors of
$\log(\epsilon_1)$. In order to fairly compare BK and SPT, we have
carefully complemented both of them with a compatible ensemble $\calD$
of the currently most precise cosmological data sets. We have not
attempted to estimate the \emph{overall} probability of $\uSPT+\calD$
to be incompatible with $\uBK+\calD$, as for instance by using the
$R$-factors. Indeed, our data analysis shows that over their $68$
dimensions both data sets differ by two-sigma along one parameter
direction only. From a Bayesian point of view, they are definitely
compatible (see Ref.~\cite{Martin:2014lra} for a similar analysis). As
such, our findings are not more than a surprising tension coming
solely from the $B$-mode data.

One may wonder why our analysis leads to a higher preference for
non-vanishing $r_\epsilon$ than the one reported in
Ref.~\cite{SPT-3G:2025vtb}. A difference resides in our use of the
more theoretically justified slow-roll primordial power spectra, as
opposed to phenomenological power laws. Another more likely
possibility is the choice of our complementary data sets $\calD$. In
order to test the robustness of our results, we have reproduced
essentially the same discrepancy between BK and SPT by either removing
the lensing or BAO data, as well as by trading the {\SROLL}
polarisation data for the legacy {\PLANCK} data ($\texttt{lowE}$). We
have also tested the same exact analysis by replacing the {\EBOSS} BAO
data by the DESI data and, even though some cosmological parameter
posteriors are accordingly shifted, the posterior on
$\log(\epsilon_1)$ remains unaffected. Various other numerical
settings in {\CAMB} have also been tested to ensure the accuracy of
the lensing computations.

\begin{figure}
  \begin{center}
    \includegraphics[width=\figw]{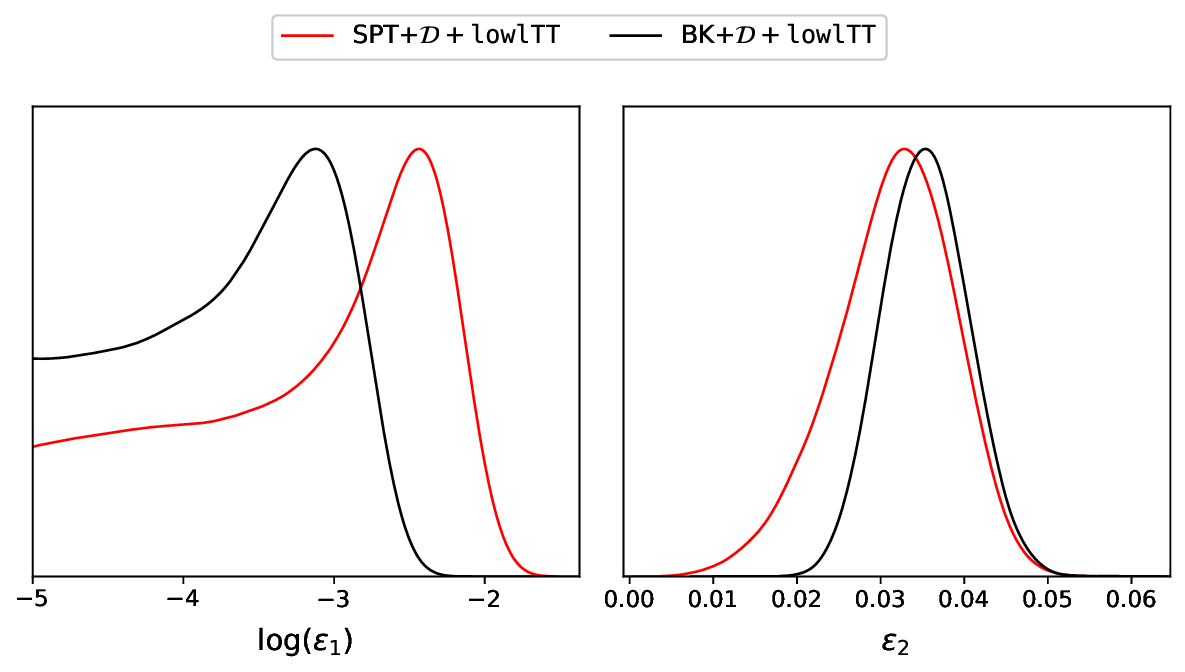}    
    \includegraphics[width=\figw]{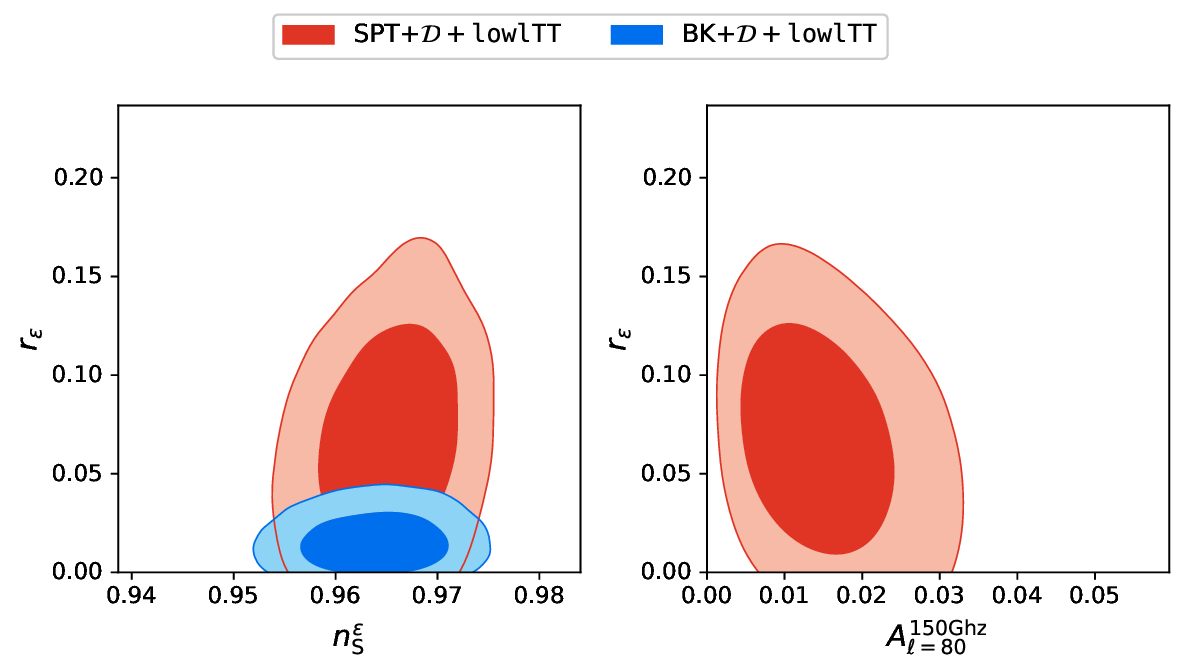}
    \caption{Effects of including the $\TT$ lowest multipoles
      from {\PLANCK} on some selected slow-roll and derived power law
      parameters (to be compared with \cref{fig:Psr1D} and
      \cref{fig:Ppl2D}).}
    \label{fig:lowltension}
  \end{center}
\end{figure}

Finally, we have also tested the inclusion of the {\LOWLTT} {\PLANCK}
CMB data which correspond to the multipoles $\ell=2$ to $\ell=29$ for
the temperature anisotropies. The {\LOWLTT} likelihood is specific to
these scales and pixel-based as to enforce an optimal component
separation~\cite{Planck:2018yye}. As discussed in the introduction,
primordial gravitational waves contribute to a small amount of the
overall $\TT$ power spectrum at the largest length scales and they
should be (weakly) constrained by the {\LOWLTT} likelihood. Various
relevant posteriors obtained from $\calD+\LOWLTT$ have been
represented in \cref{fig:lowltension}. For BK, the posterior of
$\log(\epsilon_1)$ is unchanged whereas the one associated with SPT is
less peaked and slightly shifted to lower values. As a result, there
is also a small tension between the SPT $B$-mode measurements and the
{\PLANCK} {\LOWLTT} data. This observation suggests that the $B$-modes
excess we have found here, as measured by SPT, could rather be due to
a yet to be understood
foreground~\cite{Ferreira:2025lrd}. Nonetheless, one should
probably keep an open mind to the alternative which is that the BK
data are overmarginalising in their nuisances a non-vanishing
contribution of $B$-modes of inflationary origin. In all possible
situations, our findings should provide additional motivations for new
and independent measurements of the CMB $B$-modes, possibly with new
map making methods~\cite{Keck:2025xzv}, other
telescopes~\cite{CMB-S4:2022ght}, and most importantly, elsewhere, if
not everywhere, in the sky~\cite{LiteBIRD:2022cnt}.

\acknowledgements

This work is supported by the ESA Belgian Federal PRODEX Grants
$\mathrm{N^{\circ}} 4000143201$ and $\mathrm{N^{\circ}}
4000144768$. Computations have been performed on the
\href{https://curl.group}{CURL} development cluster and the Center for
High Performance Computing and Mass Storage
(\href{https://www.uclouvain.be/en/cism/cism-platform}{CISM}) at UCLouvain.

\bibliographystyle{JHEP}
\bibliography{biblio}

\end{document}